\documentstyle[aps,prb,multicol,epsf]{revtex}

\begin{document}

\title{
Mott transition and heavy fermion state in the pyrochlore Hubbard model
}

\author{Satoshi Fujimoto}
\address{
Department of Physics,
Kyoto University, Kyoto 606-8502, Japan
}

\date{\today}
\maketitle
\begin{abstract}
We investigate the interplay between geometrical frustration and
strong electron correlation based upon the pyrochlore Hubbard model.
In the half-filling case, using the perturbative expansion in terms of
electron correlation, we show that the self-energy shows a divergent 
behavior leading the system into the Mott insulating state, 
in which quantum disordered spin liquid without magnetic long-range order
realizes.
In the hole-doped case, we obtain heavy-fermion-like Fermi liquid state.
We also calculate the neutron cross section which is well consistent
with recent neutron scattering experiments for itinerant pyrochlore systems.
\end{abstract}


\begin{multicols}{2}
\section{Introduction}

Recently, the role played by geometrical frustration in strongly correlated
electron systems has attracted renewed 
interests.\cite{shiga1,kondo,take,toku,sato,mand}
The pyrochlore lattice, namely, corner-sharing tetrahedra (FIG.1),
is one of the typical system in which the geometrical frustration
is crucial to determine its properties.
Here we investigate effects of geometrical frustration on itinerant
electron systems based upon the pyrochlore Hubbard model.
One of the purpose of this paper is to investigate
the Mott transition in the absence of magnetic long-range order.
According to the study of the Heisenberg model on the pyrochlore lattice,
it has been pointed out that the ground state of the insulating state 
is quantum disordered spin 
liquid.\cite{moe,cana,lieb,iso1,ueda,gin,koga,tsune,cana2}
The critical character of the Mott transition without magnetic order
is quite different from that accompanying antiferromagnetic one.
Moreover the character of the quantum spin liquid has not yet been unveiled
sufficiently.
It is expected that the study from electron systems 
will shed new light on this subject.

Another purpose of this paper is to reveal how the geometrical frustration
affects strong electron correlation effects in the metallic state.
Some recent experiments on the pyrochlore itinerant electron systems
have reported that remarkable heavy-fermion-like behaviors manifest 
in these systems implying the crucial role 
of the geometrical frustration.\cite{shiga1,kondo}
Here we show that the heavy-fermion state is realized 
in the pyrochlore Hubbard model in the vicinity of the
half-filling. 
We also show that the dynamical spin susceptibility obtained from our
model away from the half-filling is consistent with recent neutron
scattering experiments for itinerant pyrochlore systems.\cite{bal}

The organization of this paper is as follows.
The model and the basic method are given in Sec. II.
In Sec. III, we discuss about the Mott transition and
the realization of the quantum disordered spin liquid in the half-filling case.
In Sec. IV, we consider the hole-doped case focusing on
the heavy fermion state and its magnetic properties.

\begin{figure}
\centerline{\epsfxsize=5.5cm \epsfbox{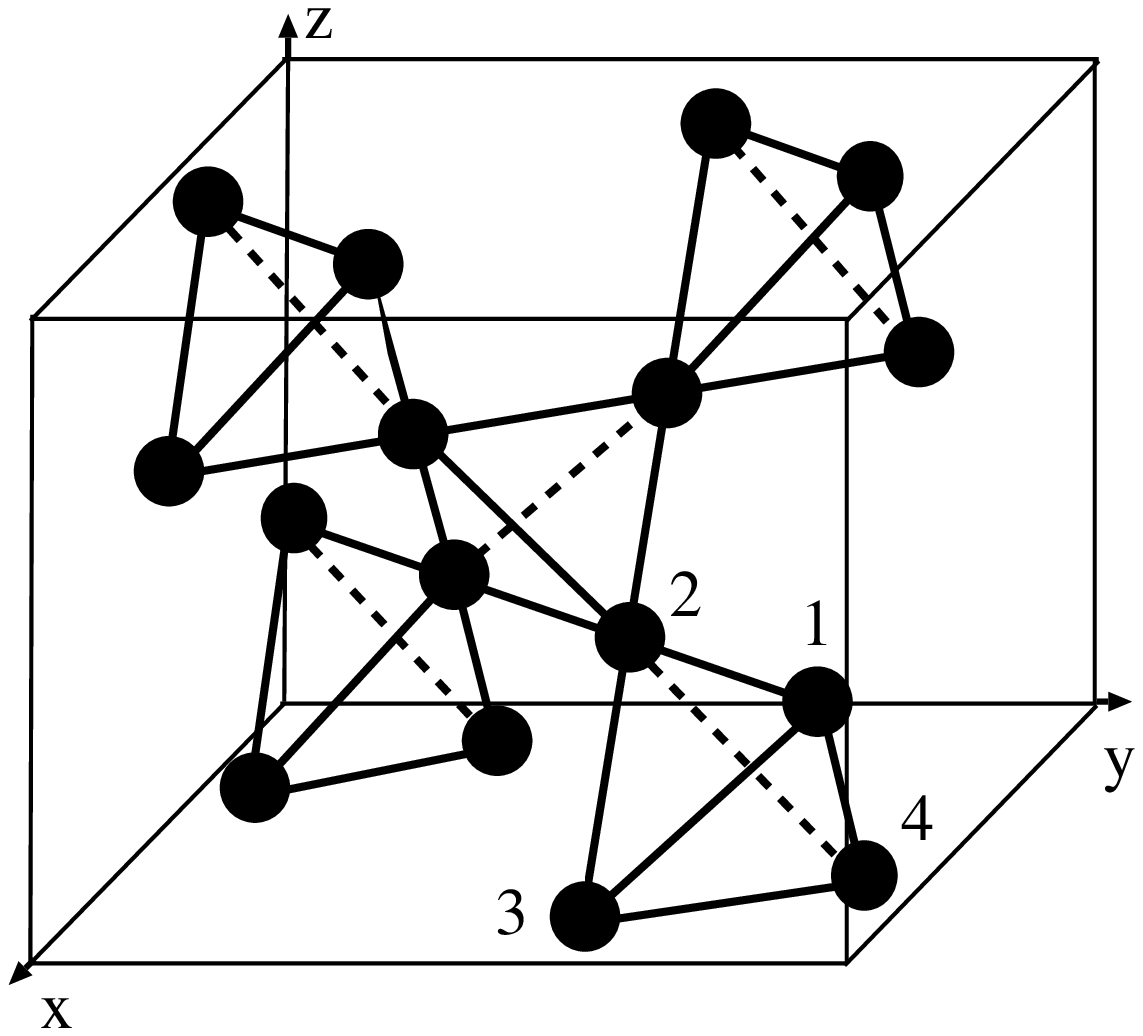}}
{FIG. 1. Pyrochlore lattice.}
\end{figure}

\section{Model Hamiltonian and method}

The Hamiltonian of the model is given by,
\end{multicols}
\begin{eqnarray}
H=-2\sum_{k,\sigma}\psi^{\dagger}_{k\sigma}
\left(
\begin{array}{cccc}
0&\cos(k_x+k_y)&\cos(k_y-k_z) &\cos(k_x-k_z)  \\
\cos(k_x+k_y)&0&\cos(k_x+k_z)&\cos(k_y+k_z) \\
\cos(k_y-k_z)&\cos(k_x+k_z)&0&\cos(k_x-k_y) \\
\cos(k_x-k_z)&\cos(k_y+k_z)&\cos(k_x-k_y)&0
\end{array}
\right)
\psi_{k\sigma}   
+U\sum_{i}\sum_{\nu=1}^4c^{\dagger}_{i\nu\uparrow}c_{i\nu\uparrow}
c^{\dagger}_{i\nu\downarrow}c_{i\nu\downarrow}, 
\end{eqnarray}
\begin{multicols}{2}
Here $c_{i\nu\sigma}(c_{i\nu\sigma}^{\dagger})$ is an annihilation 
(creation) operator of electrons with spin $\sigma$
on the $\nu$-th site of the $i$-th tetrahedron.
$\psi_{k\sigma}^{\dagger}=
(c^{\dagger}_{k,1\sigma},c^{\dagger}_{k,2\sigma},
c^{\dagger}_{k,3\sigma},c^{\dagger}_{k,4\sigma})$. 
The kinetic term is diagonalized as,
$H_{\rm kin}=\sum_{k\sigma}\sum_{\nu=1}^4E_{k\nu}
a^{\dagger}_{k\nu \sigma}a_{k\nu \sigma}$,
where $E_{k1}=E_{k2}=1$, $E_{k3}=-1+\sqrt{1+t_k}$, $E_{k4}=-1-\sqrt{1+t_k}$ 
with $t_k=\cos(2k_x)\cos(2k_y)+\cos(2k_y)\cos(2k_z)+\cos(2k_z)\cos(2k_x)$.
The basis in the diagonalized space is obtained from the canonical 
transformation,
$\psi_{k\sigma}^{\dagger}=\sum_{\nu=1}^4\vec{s}_{\nu}(k)
a_{k\nu\sigma}^{\dagger}$.
Using the abbreviation, $s(x\pm y)\equiv \sin(k_x\pm k_y)$ etc., we write down 
$\vec{s}_{1,2}$(k),
\begin{eqnarray}
\vec{s}_1(k)=(s(x+z),s(y-z),-s(x+y),0)/n_{+},
\end{eqnarray}
where $n_{\pm}=\sqrt{s(x\pm z)^2+s(y\mp z)^2+s(x+y)^2}$, and,
\begin{eqnarray}
\vec{s}_2(k)&=&(s(x+z)s(x-z)s(y-z) \nonumber \\
&&-s(y+z)(s(y-z)^2+s(x+y)^2), \nonumber \\
&& s(x+z)s(y+z)s(y-z)  \nonumber \\
&&-s(x-z)(s(x+z)^2+s(x+y)^2), \nonumber \\
&& -s(x+y)(s(x+z)s(y+z)+s(x-z)s(y-z)), \nonumber \\
&& s(x+y)n_{+}^2)/n_2,
\end{eqnarray}
for $k_x+k_y\neq 0$, where $n_2=n_{+}[n_{+}^2n_{-}^2-(s(x+z)s(y+z)
+s(x-z)s(y-z))^2]^{\frac{1}{2}}$, and,
\begin{eqnarray}
\vec{s}_2(k)&=&(-s(2x),-s(2x),2s(x-z),2s(x+z)) \nonumber \\
&\times& 1/\sqrt{2s(2x)^2+4s(x-z)^2+4s(x+z)^2}
\end{eqnarray}
for $k_x+k_y=0$.
The expressions of $\vec{s}_3(k)$ and $\vec{s}_4(k)$ are very complicated.
However in the following, we need only $\vec{s}_3(k)$ for small $k$, 
which is given by,
\begin{eqnarray}
\vec{s}_3(k)&=&(-k_x-k_y+k_z,k_x+k_y+k_z, \nonumber \\
&& -k_x+k_y-k_z,k_x-k_y-k_z)
/2\vert k\vert.
\end{eqnarray}

At the half-filling, in the absence of electron-electron interaction,
 $E_{k3}$ and $E_{k4}$ are filled completely, and the two degenerate flat bands
are empty. $E_{k3}$ touches with the two flat bands at the $\Gamma$ point
of the Brillouin zone.
This state is a gapless band insulator. 
We now consider the effect of electron interaction on this state.
Since, in the case of the half-filling or in the vicinity of the half-filling,
the band $E_{k4}$ is sufficiently far from the Fermi level,
it does not affect low energy properties and is negligible
in the following argument.
Then, the single-particle Green's functions of $a_{k\nu\sigma}$-electron are 
determined by the following equation.
\begin{eqnarray}
\sum_{\lambda=1}^3[(\varepsilon+\mu-E_{k\lambda})\delta_{\mu\lambda}
-\Sigma_k^{(\mu\lambda)}(\varepsilon)]
G^{(\lambda\nu)}_k(\varepsilon)=\delta_{\mu\nu}.
\end{eqnarray}
We put the chemical potential $\mu=1$.
In general, the off-diagonal self-energy $\Sigma^{(\mu\nu)}$($\mu\neq\nu$)
is not negligible.
However, in the vicinity of the $\Gamma$ point, at which 
the most important scattering processes occur, the off-diagonal terms
vanish because of the momentum dependence of $\vec{s}_{\nu}(k)$.
Thus in the following argument, we neglect the off-diagonal self-energy.

\section{The half-filling case}

\subsection{Self-energy}

To investigate the electronic state at the half-filling, 
we calculate the self-energy. 
Recently, Isoda and Mori obtained the diagonal self-energy of this model 
up to the second order in $U$,\cite{iso2} which is given by, 
${\rm Re}\Sigma(\varepsilon)\sim \sqrt{-\varepsilon}$
for $\varepsilon<0$.
However, as will be shown here, the higher order corrections to
the self-energy give rise more singular contributions and a drastic change 
of the electronic state, namely, the Mott transition.

\begin{figure}
\centerline{\epsfxsize=6cm \epsfbox{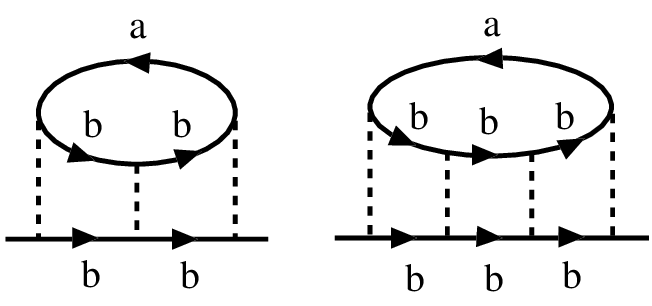}}
{FIG. 2. The most singular diagrams of the self-energy up to the forth-order
in $U$. $a$ denotes $a_3$-electron. $b$ denotes $a_1$ or $a_2$-electron.
The dotted line denotes the interaction $U$.}
\end{figure}

At the half-filling, the $E_{k3}$-band is completely filled,
while the flat bands are empty.
Hence, at zero temperature,
particle-particle channel between $a_3$-electron and
$a_1$ or $a_2$-electron vanishes.
On the other hand, particle-hole channel gives non-vanishing 
contributions only for the pair of $a_3$-electron and 
$a_1$ or $a_2$-electron.
Then the most singular diagrams up to the forth order in $U$ 
are those shown in FIG.2.
Since the Fermi level is at the position between the two flat bands and
$E_{k3}$-band, we shift the chemical potential infinitesimally, 
$\mu\rightarrow 1-\delta$. After integrating over energy and momentum,
we take the limit $\delta\rightarrow 0$.
This procedure is required in this singular perturbative calculation
to fix the density of electrons $n=1$.
Neglecting the momentum dependence of $\vec{s}_{\nu}(k)$ does not
affect the leading singular behaviors.
Then, we can carry out the calculation analytically.\cite{agd}
For $\varepsilon<0$, we obtain,
\begin{eqnarray}
{\rm Re}\Sigma(\varepsilon)\sim\mbox{const.}-cU^2\sqrt{|\varepsilon |}
-c\frac{U^3}{2\sqrt{|\varepsilon |}}
+c\frac{U^4}{4|\varepsilon |^{\frac{3}{2}}},
\label{reself}
\end{eqnarray}
where $c$ is a positive constant of order unity.
The real part of the self-energy for hole-like excitations is 
divergent.
The same divergent behavior appears in the imaginary part, however, for
particle-like excitations.
Higher order corrections also show stronger divergence.
This divergent behavior means that the unperturbed state is unstable
in the presence of electron correlation and that the single particle energy
gap is generated.
To see this we take only up to the third order term of the self-energy.
Hole-like excitations are possible only for $E_{k3}$-band 
in the half-filling case at zero temperature.
Thus the single particle energy of this band measured from the Fermi level
is changed to $\varepsilon_k\sim -U^2\equiv -\Delta$ for $k=0$.
$E_{k3}$ band is pushed down to the lower energy, and
the energy gap opens between the flat bands and $E_{k3}$-band.
This gap generation due to electron correlation signifies that
the systems is in the Mott insulating state.
Although the exact magnitude of the gap should be determined by taking 
into account higher order corrections in $U$,
the gapful state is the self-consistent solution for non-vanishing $U$.
Thus an infinitesimally small $U$ drives the system into the Mott insulator
at the half-filling.
In the ground state, the two bands below the gap, 
$E_{k3}$, $E_{k4}$,
are filled with up and down spins.
Hence the system is in the spin singlet state.
It should be stressed that, after the mass gap generation, 
the singularities of the perturbative expansion are eliminated, and
as a result, the gapful state is stabilized even if one take into account
higher order corrections, which just renormalize the magnitude of the gap.

\subsection{Quantum spin liquid state}

To see the magnetic property of the Mott insulating state obtained 
in the previous section, we compute
spin-spin correlation functions, 
$\chi_{\mu\nu}(q,\omega)=
\langle S_{\mu}^{+}(q,\omega)S_{\nu}^{-}(-q,-\omega)\rangle$,
where $S_{\nu}^{+}=\sum_k c^{\dagger}_{k+q\nu\uparrow}c_{k\nu\downarrow}$.
We carry out perturbative expansion in terms of $U$ using
the Green's function of the gapful state,
$G^{(33)}(\varepsilon)\sim [\varepsilon-E_{k3}+\Delta]^{-1}$, 
as the unperturbed propagator.
Calculating the diagrams up to the third order in $U$,
we have, ${\rm Im}\chi(q,\omega)\sim \sqrt{\omega-\Delta}\Theta(\omega-\Delta)$
with $\Theta(x)$ a step function.
The spin gap exists between the spin singlet ground state and 
the triplet state. Thus the system is in the quantum disordered spin 
liquid state without magnetic long-range order.
The result may not be changed qualitatively by higher order corrections.  
These results are consistent with the previous studies for the pyrochlore
Heisenberg model.\cite{cana,gin,koga,tsune,cana2}
The singularity of the self-energy (\ref{reself}) is deeply
related with the accidental degeneracy at the $\Gamma$ point in the
momentum space, which may cause an instability of lattice structure.
This property may be relevant to the metal-insulator transition
of ${\rm Tl_2Ru_2O_7}$, which accompanies a structure change.\cite{take,ogu}

\section{The hole-doled case}

\subsection{Heavy fermion state}

In the hole-doped case, 
the singularities appeared in the half-filling case
are eliminated by the presence of the cutoff which is 
the chemical potential $\mu$ measured from the half-filling level.   
Then the Fermi liquid metallic state is realized.
However, in the vicinity of the half-filling, the effective mass is much
enhanced by electron correlation.
Up to the third order in $U$, the leading term of the mass enhancement 
factor $z_k$ is,
\begin{eqnarray}
z_k^{-1}\equiv 1-\left.\frac{\partial Re\Sigma(\varepsilon)}
{\partial\varepsilon}
\right|_{\varepsilon\rightarrow 0}
\sim \frac{U^2}{\sqrt{\vert \mu\vert}}
+\frac{U^3}{\vert\mu\vert^{\frac{3}{2}}}.
\label{reself2}
\end{eqnarray}
As the electron filling approaches the half-filling value
$\vert\mu\vert\rightarrow 0$,
the mass enhancement factor shows the divergent behavior indicating
the precursor of the Mott transition.
Such a large mass enhancement is actually observed in the specific heat
measurement for
some pyrochlore itinerant systems like ${\rm Y(Sc)Mn_2}$.\cite{shiga1}

\begin{figure}
\centerline{\epsfxsize=5.5cm \epsfbox{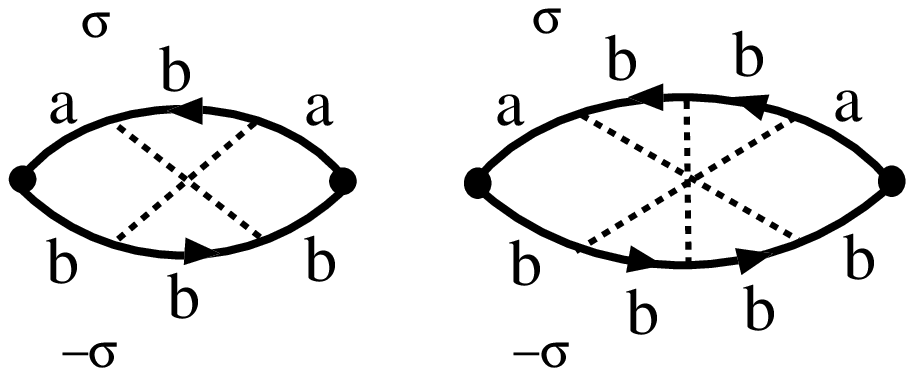}}
{FIG. 3.The most singular diagrams of spin-spin correlation functions 
up to the third-order in $U$. }
\end{figure}

\begin{figure}
\centerline{\epsfxsize=6.5cm \epsfbox{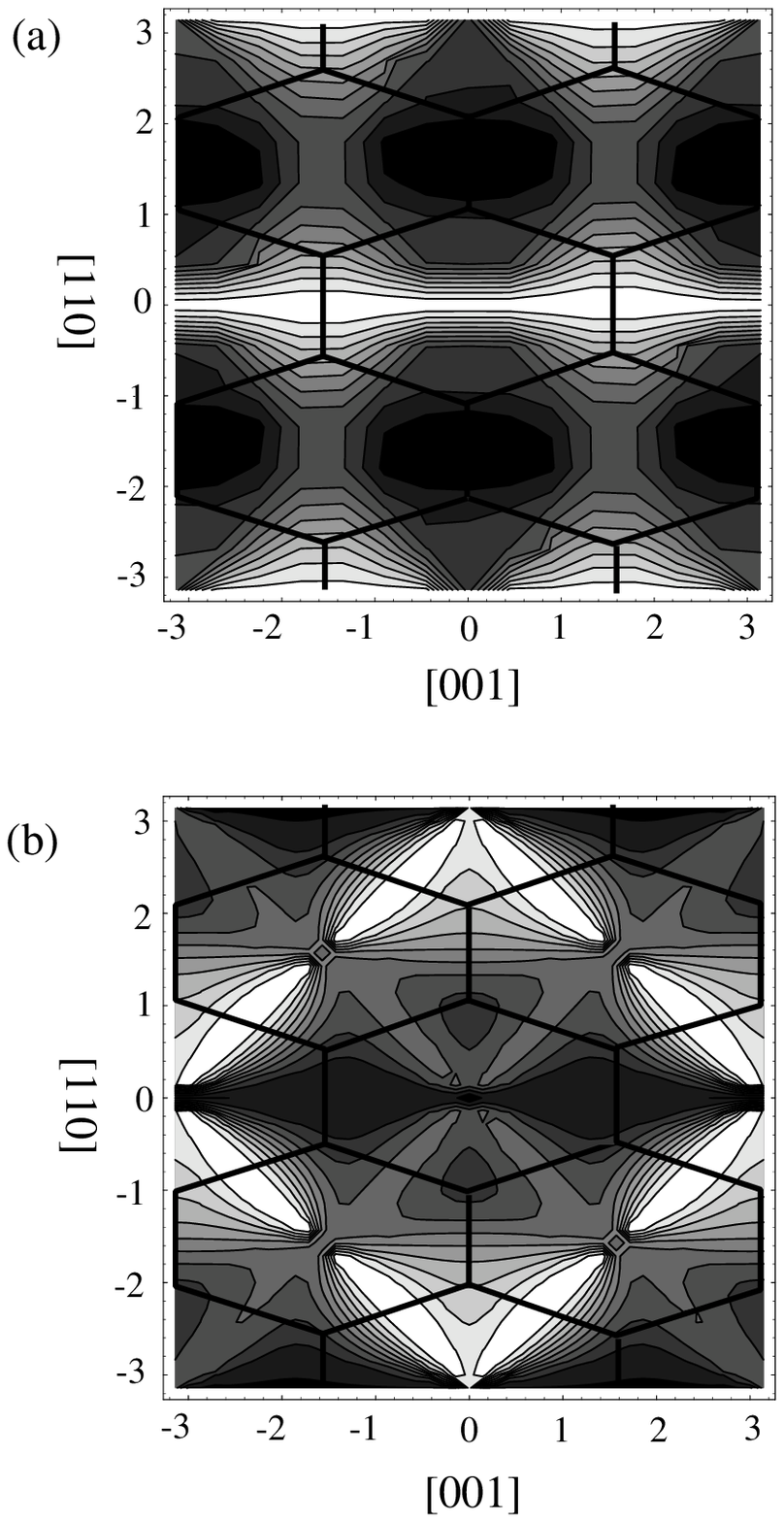}}
{FIG. 4. (a) The structure factor $g_{12}(q)$ plotted in the [001]-[110]
plane. The black line denotes the zone boundary. The brighter regions
have the stronger intensity. 
(b) The neutron cross section.}
\end{figure}

\subsection{Magnetic properties and the neutron cross section}

The largely enhanced effective mass obtained in the previous section
is a remarkable property
of geometrically frustrated electron systems in which
several modes of spin fluctuations compete with each other, and
magnetic long-range order is suppressed.
The spin fluctuation is almost localized in real space 
as in the case of $f$-electron based heavy fermion systems.
This property is seen more clearly in the momentum dependence of
the spin-spin correlation function.
We calculate it by the perturbative calculation.
Since the magnetic frustration is so strong, 
any approximations which neglect the coupling between
several modes of spin fluctuation like random phase approximation(RPA) or
fluctuation exchange approximation(FLEX) are invalid for this system.
Actually the most singular contributions up to the order of $O(U^3)$
comes from the diagrams shown in FIG.3. 
To simplify the calculation we exploit the following approximation.
Near the half-filling, the momenta running in $G^{(33)}$-line
take small values in the vicinity of the Fermi surface.
Moreover for small $k$, $\vec{s}_{3}(k)$ does not depend on $\vert k\vert$.
Thus we can separate the integral over $\vert k\vert$ and the angle
for the momenta running in $G^{(33)}$-line.
Then, the spin-spin correlation function is decomposed into
the momentum part and the energy part,
$\chi_{\mu\nu}(q,\omega)=f(\omega)g_{\mu\nu}(q)$.
Such a factorization of the spin-spin correlation function is indeed
observed in an inelastic neutron scattering experiment 
for ${\rm Y(Sc)Mn_2}$.\cite{bal}
The energy part is given by,
\begin{eqnarray}
f(\omega)\sim -a_1U^2\log\frac{\vert \omega+2\mu\vert}{E_c}
-a_2 U^3\frac{1}{\omega+2\mu}+...
\end{eqnarray}
for small $\omega$. Here $a_1$, $a_2$ are constants.
Thus in the vicinity of the half-filling,
spin fluctuation is much enhanced.
The presence of the giant spin fluctuation is also consistent with
several experiments for itinerant pyrochlore systems.\cite{shiga1,bal}
It is noted that such a large enhancement of $\chi_{\mu\nu}(q,\omega)$
is not obtained by RPA or FLEX approximations for the appropriate value
of $U$.
We calculate the structure factor $g_{\mu\nu}(q)$ numerically.
In FIG.4(a), we plot the intensity of $g_{12}(q)$
in the [110]-[001] plane of the reciprocal lattice.
$g_{12}(q)$ is almost constant in the [001]-direction,
indicating that the spin fluctuation is strongly localized in this direction.
It also has a small peak on the $q_x=q_y=0$ line
showing the presence of small fluctuation toward a collinear magnetic
order, as was pointed out in the study of the Heisenberg model.\cite{cana}
However this mode competes with the other modes which exist in $g_{13}(q)$,
$g_{23}(q)$, $g_{14}(q)$, $g_{24}(q)$, and thus
such a magnetic order is suppressed. 

In FIG.4(b), we show the calculated result of the neutron cross section,
$\sum_{\mu,\nu} e^{-iG(R_{\mu}
-R_{\nu})}\chi_{\mu\nu}(q,\omega)$, in the [110]-[001] plane.
Here $G$ is a reciprocal lattice vector.
The positions of the maximum are almost
consistent with the neutron scattering experiment 
for ${\rm Y(Sc)Mn_2}$.\cite{bal} Note that this structure is very similar to
that found for the Heisenberg model.\cite{cana}
Thus the itineracy of electron affect little 
the structure in the momentum space, though the dynamical properties between
the Mott insulating state and the metallic state are quite different;
namely, the former is gapful and the latter is gapless.
The momentum dependence of $\chi_{\mu\nu}(q,\omega)$ implies
that the spin fluctuation is localized in a tetrahedron forming
collective 4-spin singlet.\cite{iso1,cana}
Such a localized character of spin fluctuation is very similar to
magnetic properties of $f$-electron based heavy fermion systems.
However, in the latter systems, magnetic long-range order is suppressed by
the presence of Kondo temperature higher than magnetic ordering temperature.
In our system, the geometrical frustration is crucial. 

\subsection{Compressibility}

Here we discuss the charge response of the heavy-fermion state
near the half-filling.
In general, the charge susceptibility $\chi_c$ 
in the vicinity of the Mott metal-insulator
transition point shows quite different behaviors depending on
the density of states near the Fermi level.\cite{ima}
In our systems, the bare density of states in the hole-doped case is 
non-singular, though four-point vertices are much enhanced by 
electron correlation.
The perturbative calculation of the charge susceptibility gives 
singular terms in the limit of $\mu\rightarrow 0$, namely,
$\chi_c\sim C_1U^2\log\vert\mu\vert+C_2U^3/\vert\mu\vert+\cdot\cdot\cdot$.
Thus it is highly non-trivial how the charge susceptibility behaves,
as the electron filling approaches the half-filling value.
To see this we resum the most singular terms of 
the perturbative expansion in $U$.
As was done before, we neglect the momentum dependence of $\vec{s}_{\nu}$.
We first consider the terms which contain only one irreducible four-point
vertex.
The most singular terms of this type are diagrammatically expressed as FIG.5(a)
These terms are calculated as,
\begin{eqnarray}
\chi^{\rm irr}_c=C_2U^2\log\vert\mu\vert+C_3U^3/\vert\mu\vert
+C_4U^4/\vert\mu\vert^2+\cdot\cdot\cdot.
\end{eqnarray}
Here $C_2$,$C_3$, and $C_4$ are positive constants.
On the other hand, the term which contains $n$ irreducible 
four-point vertices shown in FIG.5(b) gives the contribution 
$\sim [-c\chi^{\rm irr}_c]^n/\vert\mu\vert^{\frac{n-1}{2}}$ with $c$ 
a positive constant.
Summing up all most singular terms, we have,
\begin{eqnarray}
\chi_c=a\sqrt{\vert\mu\vert}-\frac{\chi^{\rm irr}_c}
{1+\frac{c}{\sqrt{\vert\mu\vert}}\chi^{\rm irr}_c}.
\end{eqnarray} 
Thus as the electron filling approaches the half-filling value 
$\mu\rightarrow 0$, the charge susceptibility decreases toward zero,
$\chi_c\rightarrow 0$, indicating that the system becomes incompressible.

\begin{figure}
\centerline{\epsfxsize=8cm \epsfbox{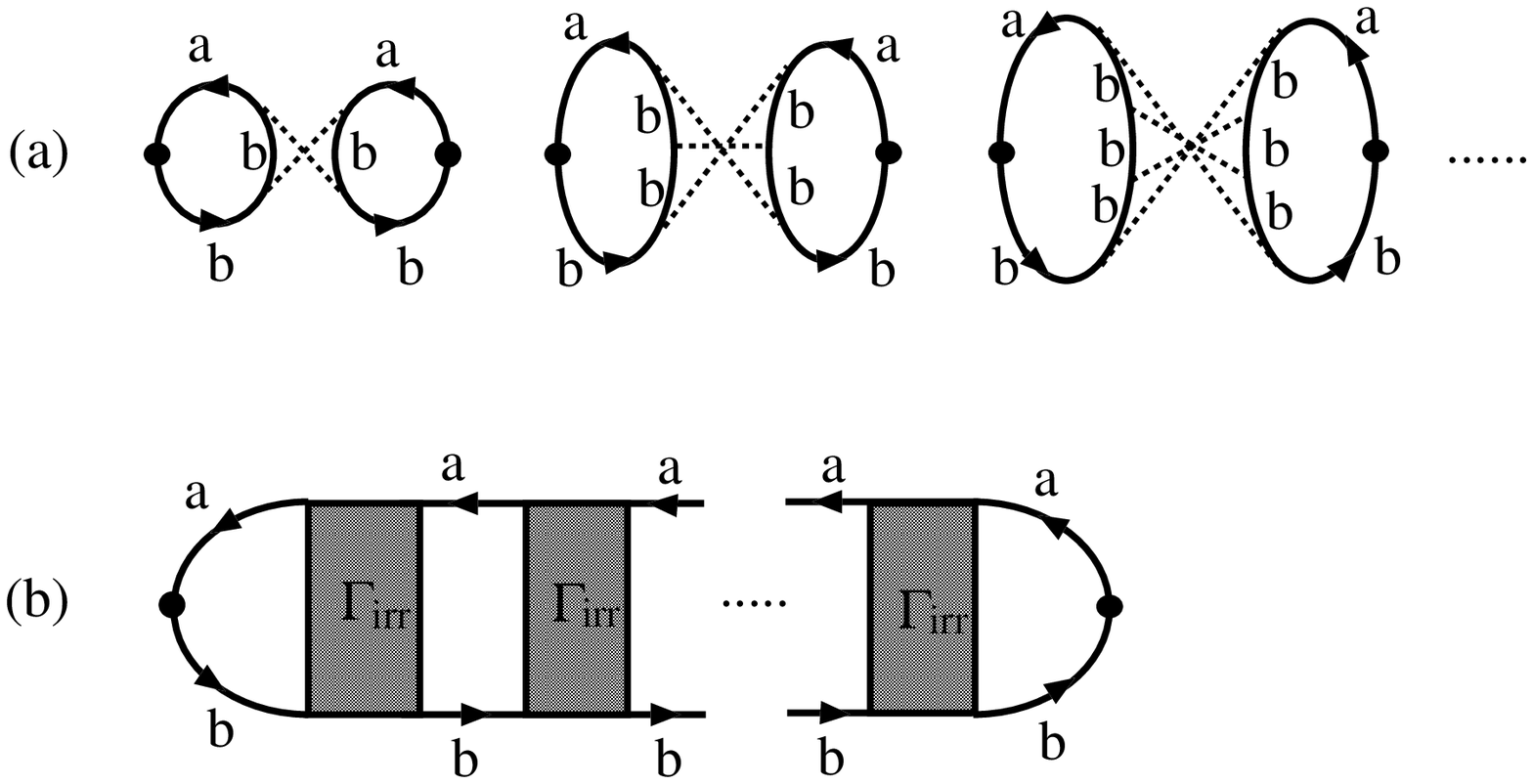}}
{FIG. 5. Diagrams of the charge susceptibility. $\Gamma_{irr}$ 
is the irreducible 4-point vertex.}
\end{figure}

\section{Summary}

In this paper,
we investigate the Mott transition, strong correlation effects
and spin dynamics of the pyrochlore Hubbard model.
We show that, at the half-filling, 
electron-electron interaction leads the system into
the Mott insulator in which 
the quantum disordered spin liquid state realizes. 
In the hole-doped case, we show that the effective mass is anomalously
enhanced in the vicinity of the half-filling signifying
the heavy fermion state.
We have also calculated the neutron cross sectin of this state 
which is consistent with the recent experiment for
some itinerant pyrochlore systems like ${\rm Y(Sc)Mn_2}$.

Although we carried out the perturbative calculation up to the
fourth order in $U$, the singular divergent behavior of the self-energy
shown in Sec. III implies that we need to resum all order singular
diagrams. Since the most singular diagrams have the specific structure
as discussed in Sec. III, we can carry out this program in principle.
We would like to address this issue in the near future.

\acknowledgements{}
The author would like to thank K. Yamada, K. Ueda, N. Kawakami, 
H. Tsunetsugu, H. Ikeda for invaluable discussions.
This work was partly supported by a Grant-in-Aid from the Ministry
of Education, Science, and Culture, Japan.

\end{multicols}
                                                                    
\end{document}